# LesionStudio: A Patient-Informed Platform for Task-Specific Optimization of Emission Tomography Reconstruction and Analysis

Narges Aghakhanolia[1,2,*], James Fowler[2,3], Ahmad Shariftabrizi[4], Omid Gharibi[2,3], Behnam Panahi[1], Sinan Akay[4], Arman Rahmim[1,2,3,5], Pedro L. Esquinas[5,6]

*1. School of Biomedical Engineering, University of British Columbia, Vancouver, BC, Canada, V6T 1Z3*

*2. Department of Basic and Translational Research, BC Cancer Research Institute, Vancouver, BC, Canada, V5Z 1L3*

*3. Department of Physics and Astronomy, University of British Columbia, Vancouver, BC, Canada, V6T 1Z1*

*4. Division of Nuclear Medicine, Department of Radiology, University of Iowa Carver College of Medicine, Iowa City, IA, USA, 52242*

*5. Department of Radiology, University of British Columbia, Vancouver, BC, Canada, V5Z 1M9*

*6. Department of Molecular Imaging and Therapy, BC Cancer, Vancouver, BC, Canada, V5Z 1L3*

*** Corresponding Author:** narges.aolia@gmail.com

## Abstract

**Background:** Optimizing PET and SPECT reconstruction for quantification and lesion-detection tasks is limited by conventional phantoms, which do not capture patient-specific background heterogeneity, realistic lesion morphology and texture, or lesion-background interactions.

**Objective:** To develop and evaluate LesionStudio, an open-source platform for patient-informed, task-specific optimization of PET and SPECT reconstructions using synthetic lesions inserted into real clinical acquisitions.

**Methods:** LesionStudio, a Python framework built on PyTomography, inserts lesions into PET/SPECT data using modality-specific forward projection and stochastic background ablation. Clinical realism was evaluated by blinded physician review of 30 inserted lesions in eight DLBCL patients from a 1,579-lesion FDG lymphoma library. Quantification and detectability were assessed in FDG PET using spherical and real-patient-derived lesions varying in size, LBR, and site, reconstructed with OSEM and BSREM.

**Results:** Inserted lesions were visually indistinguishable from true disease (AUC = 0.49). OSEM improved the patient-derived quantification limit from ~3.7 mL at iteration 4 to ~2.0 mL at iteration 10, but with increased noise and edge artifacts. Weakly regularized BSREM ($\beta = 5$) achieved a similar ~2.0 mL limit with greater stability, whereas stronger regularization increased under-recovery. Detectability was strongly contrast-dependent: no lesions were detected at LBR 2:1, whereas the minimum detectable volume ranged from 0.70 to 3.63 mL at LBR 4:1 and decreased to 0.70 mL at LBR 6:1–8:1 for both OSEM and weakly regularized BSREM.

**Conclusion:** Reconstruction settings can shift SUV metrics and detectability enough to affect interpretation, so no single protocol suits every task. OSEM required early stopping for detection and feature stability but more iterations for mean-SUV quantification, whereas weakly regularized BSREM matched both at one setting.

## 1. Introduction

Positron Emission Tomography (PET) and Single Photon Emission Computed Tomography (SPECT) are critical *in vivo* molecular imaging modalities in the management of a wide range of oncologic and nononcologic diseases [1,2]. By employing radiopharmaceuticals, these techniques enable visualization and quantification of metabolic and biochemical processes, thereby supporting early tumor detection, disease staging, treatment planning, therapy monitoring, and post-treatment dosimetry [3,4].

To fully leverage the diagnostic and therapeutic potential of PET and SPECT, careful optimization of image reconstruction algorithms, reconstruction parameters, and post-reconstruction filtering is essential [5,6]. The quality of the reconstructed images and, consequently, the accuracy of lesion detection and radiopharmaceutical uptake quantification depend strongly on these choices [7].

However, optimizing reconstruction methods for detection and quantification tasks remains challenging. Traditional approaches rely heavily on both physical and digital phantoms [8,9], which allow explicit design of lesion parameters for targeted evaluation. Yet, these models often fall short of replicating the complexity of clinical imaging, particularly due to heterogeneous background activity and anatomical variability in real patient data. For instance, partial-volume effects in organ-shaped geometries differ substantially from those in simple spherical phantoms [10], and recovery coefficients vary with lesion location within the field of view and surrounding background uptake [11]. Consequently, reconstruction parameter optimization based solely on phantoms can yield suboptimal results in real patient data. While Monte Carlo simulations are also employed to model photon transport and detector physics in emission tomography [12], their use in clinically realistic optimization is constrained by computational demands and simplifying assumptions about anatomy, activity distribution, and system response. These limitations create a persistent gap between controlled experimental validation and clinical realism. This gap is especially critical when lesions occur near high-uptake structures or within complex anatomy, such as lymphoma lesions adjacent to the liver, spleen, or mediastinum in FDG PET, or lesions near the kidneys and bone in Lu-177 PSMA SPECT dosimetry [13].

To address these challenges, synthetic lesion insertion into real patient data has emerged as a robust alternative. This approach enables the introduction of synthetic lesions with precisely controlled characteristics, including size, shape, location, and uptake, while ensuring realistic system modeling [14]. Prior tools, however, have generally been developed for narrowly defined tasks, such as segmentation evaluation [15], or attenuation-correction development [16], and are typically PET-specific and MATLAB-based. This leaves a need for an openly available platform that operates directly on raw acquisition data and supports both quantification and detection optimization within a single reproducible pipeline.

To support this task-based evaluation paradigm, we developed and publicly shared LesionStudio, an open-source framework for controlled synthetic lesion insertion and image reconstruction in clinical PET and SPECT studies. The framework enables lesions with predefined size, uptake, morphology, and anatomical location to be inserted into patient data while preserving the underlying background activity distribution and noise properties of real clinical acquisitions. By operating directly on raw acquisition data and interfacing with advanced reconstruction algorithms, it provides a unified and reproducible environment for systematic assessment of lesion detectability and quantitative accuracy. We apply this framework to FDG PET in diffuse large B-cell (DLBCL) lymphoma, a disease where SUV-based quantification and lesion detection directly guide treatment decisions, to provide ground-truth-validated and clinically actionable guidance on reconstruction algorithm and parameter selection. Specifically, we evaluate how algorithm choice, iteration number, lesion size, contrast, and anatomical background jointly determine quantification accuracy and lesion conspicuity, with the aim of informing task-specific reconstruction protocols for lymphoma PET and, more broadly, for oncologic emission tomography.

## 2. Materials and Methods

### 2.1 Framework Architecture and Workflow

Inspired by the MATLAB-based PET lesion insertion method of Gabrani-Juma et al. [14], we developed LesionStudio, an interactive Python framework that advances controlled lesion insertion in clinical emission tomography. Powered by the open-source PyTomography reconstruction engine [17], it supports both PET and SPECT and integrates multimodal image registration, interactive lesion positioning, and tomographic reconstruction into a single graphical workflow for generating realistic ground-truth datasets. The overall lesion-insertion workflow is shown in Fig. 1.

The framework inserts synthetic lesions into clinical PET/SPECT acquisition data while preserving system physics, patient background activity, and noise properties. For PET, image-space lesions are forward-projected using a TOF-aware system matrix with Gaussian PSF modeling and merged with measured GE Discovery MI list-mode data. Detector geometry, attenuation and sensitivity weights, randoms, and scatter are obtained from GE lmDuetto correction files (version 02.19) and incorporated into PyTomography for consistency between simulation and reconstruction. For dual-head SPECT systems, lesions are forward-projected into clinical projection data using CT-derived attenuation maps, collimator-dependent PSF modeling, and scatter estimates from available energy windows; an optional SIMIND-based [18] hybrid Monte Carlo projector is also supported for more detailed photon-transport and collimator-detector modeling.

LesionStudio supports controlled spherical and patient-derived lesion templates. Spherical lesions enable precise control of size, position, and uptake, whereas patient-derived templates provide realistic morphology and optional intralesional heterogeneity. For this study, 1,579 segmented FDG lymphoma lesions were stored as cropped 3D templates with metadata on anatomical region, volume, and first-order radiomic features for lesion selection, curation, and matched reconstruction comparisons. The publicly shared library can be extended to other diseases or radiopharmaceuticals, and templates can be rescaled to target size and uptake while preserving heterogeneous uptake or assigning uniform intensity.

To avoid artificial signal summation in regions with physiological uptake, stochastic background ablation is applied before lesion addition. The activity contribution from the target region is forward-projected and used to estimate a deletion probability relative to the total projected signal. In PET, this probability is applied event-wise to thin the measured list-mode data; in SPECT, it is applied bin-wise to remove counts from photopeak projections using a binomial model. Thus, the inserted lesion replaces the local background signal while remaining consistent with the acquisition statistics.

After insertion, the combined data are reconstructed in PyTomography with configurable algorithms and parameters, including Ordered Subsets Expectation Maximization (OSEM) [19], Block Sequential Regularized Expectation Maximization (BSREM) [20], Ordered Subset Maximum A Posteriori One-Step Late (OSMAPOSL) [21], and Kernel Expectation Maximization (KEM) [22], with adjustable iterations, subsets, and regularization settings. The framework then automatically compares reconstructed images with known lesion ground truth across reconstruction scenarios. The graphical user interface of the framework is shown in Fig. 2. The complete framework and lesion library are publicly available at *https://github.com/qurit/LesionStudio*.

## 2.2 Experimental Design and Evaluation Methods

We evaluated the framework along two axes: diagnostic realism, defined as visual indistinguishability from true pathology, and ground-truth-based characterization of reconstruction-dependent quantification and detection performance. In this paper, we focus on PET imaging, with all experiments performed using lymphoma patients imaged with FDG PET on a GE Discovery MI-5 scanner. SPECT experiments are presented in the Supplementary Materials as a complementary validation of the framework. (Fig. S1– Fig. S3)

### *2.2.1 Diagnostic Realism*

To evaluate whether inserted lesions visually mimicked true disease, we conducted a blinded observer study using PET images reconstructed with the routine clinical setting. Thirty synthetic lesions were inserted into PET datasets from eight patients with diffuse large B-cell lymphoma (DLBCL), sampled from the patient-derived library to preserve realistic volume distributions, complex geometries, and heterogeneous uptake patterns. An experienced nuclear medicine physician reviewed images containing both native and inserted lesions, identified suspicious focal uptake, and assigned each detected lesion a five-point origin-confidence score, where 1 = *definitively real* and 5 = *definitively synthetic*. Observer-study results are reported in Section 3.1.

### *2.2.2 Quantitative Performance*

Quantitative performance was assessed against known inserted ground truth across three complementary aspects: mean-activity recovery, boundary-related edge artifacts, and the bias–noise trade-off. Recovery and bias were evaluated across lesion size, contrast, anatomical background, reconstruction algorithm, and iteration number to derive the limit of quantification (LOQ). Edge analysis assessed reconstruction-induced overshoot and undershoot, while noise–bias analysis identified operating regions balancing improved recovery against noise amplification.

**Recovery Analysis:** Quantitative recovery was evaluated using spherical and patient-derived lesions inserted into lung, bone, and liver backgrounds, representing low-, intermediate-, and high-uptake anatomical environments relevant to lymphoma imaging. Spherical lesions with diameters of 8–24 mm (0.40–7.32 mL) were inserted at LBRs of 2:1, 4:1, 6:1, and 8:1. Patient-derived lesion templates were inserted under approximately matched conditions (0.34–7.47 mL) with fixed uniform intensity to isolate the effect of irregular morphology from intralesional uptake heterogeneity. Images were reconstructed with OSEM (8 subsets, 1–10 iterations) and BSREM (8 subsets, 1–20 iterations; $\beta = 5, 10, 15, 20$, $\gamma = 2$), allowing recovery to be assessed across lesion size, contrast, anatomical background, morphology, algorithm, and convergence. Recovery was quantified using the volumetric recovery coefficient, defined as the ratio of reconstructed mean lesion uptake to inserted ground-truth uptake; percentage bias was derived from its deviation from unity [23]. Results are reported in Test 1 of Section 3.2.

**Edge Effects Analysis:** Recovery bias reflects the balance between spill-out and spill-in, but PSF-based resolution recovery and regularized iterative reconstruction can also introduce Gibbs ringing at high-contrast lesion boundaries, producing inner-edge overshoot and adjacent outer-edge undershoot [24,25,26]. To isolate reconstruction-induced ringing from intrinsic tumor uptake heterogeneity, boundary-level analysis was applied to volume-matched uniform spheres and patient-derived lesions.

For each reconstruction, a signed distance map was computed from the ground-truth lesion boundary using anisotropic voxel spacing. Inner-rim, outer-rim, and far-background shells were defined, with rim thickness set to the larger of half the reconstruction PSF FWHM and the minimum voxel dimension, and the far-background shell placed beyond at least one FWHM from the boundary. The inner-rim maximum $A^{+}$, outer-rim minimum $A^{-}$, and far-background mean $b$ were extracted. Voxel-level extrema were used rather than radially averaged profiles because tangential averaging on clinical PET grids may suppress rim oscillations and obscure the bright-rim signature [25].

Three edge-artifact metrics were computed, where $L$ denotes the inserted ground-truth peak activity. Overshoot quantified peak inflation above ground truth [25,26]:

$$\text{Overshoot}\,\% = \left(\frac{\text{SUV}_{\text{max}}}{L} - 1\right) \times 100\%$$

Undershoot quantified the low-activity dip adjacent to the lesion boundary, normalized to the far-background mean [24,25]:

$$\text{Undershoot}\,\% = \max\left(0, \frac{b - A^{-}}{b}\right) \times 100\%$$

Gibbs Activity (GA) quantified the Michelson-contrast amplitude of the boundary oscillation [25,27]:

$$\text{GA}\,\% = \frac{A^{+} - A^{-}}{A^{+} + A^{-}} \times 100\%$$

Overshoot was reported for all lesions, whereas undershoot and GA were restricted to lesions with rasterized equivalent diameter $2r_{\text{eq}} \geq \max(2\ \text{FWHM}, 2\Delta x)$, since sub-resolution lesions lack resolvable rim shells [25,28]. For patient-derived lesions, the inserted template peak served as $L$ to separate reconstruction-induced ringing from intrinsic tumor heterogeneity; for uniform spheres, positive overshoot or above-baseline GA directly reflected ringing [28].

Because GA includes a contrast-dependent baseline even without ringing, baselines of 33.3%, 60.0%, 71.4%, and 77.8% were used for LBRs of 2:1, 4:1, 6:1, and 8:1, respectively. GA values above these baselines indicated ringing-enhanced boundary oscillation, whereas lower values indicated edge blurring. Results are reported in Test 2 of Section 3.2.

**Noise-Bias Analysis:** Tests 1 and 2 characterized uptake recovery and boundary ringing, but both considered bias in isolation. To assess the balance between bias reduction and image noise, we constructed bias–noise trajectories for lesions reconstructed with OSEM and BSREM across anatomical site, lesion size, LBR, and iteration number. Bias was calculated relative to the known inserted activity, while noise was quantified as the coefficient of variation in the reference background region [29]. Results are reported in Test 3 of Section 3.2.

### *2.2.3 Detection Performance*

Having established quantitative reliability, we next evaluated lesion conspicuity using patient-derived templates inserted into lung, bone, and liver backgrounds at LBRs of 2:1, 4:1, 6:1 and 8:1. Images were reconstructed with OSEM and BSREM over 1–20 iterations. In contrast to the quantification analysis, which used fixed-intensity templates to isolate morphology-dependent recovery, here the templates retained their irregular morphology and heterogeneous intra-lesional uptake, providing a more clinically realistic assessment of conspicuity.

**Contrast-to-Noise Ratio:** Lesion conspicuity was quantified using the contrast-to-noise ratio (CNR), defined as the difference between mean lesion and background uptake normalized by background noise. CNR was evaluated after Gaussian post-reconstruction filtering with FWHM values of 1–8 mm to assess how smoothing affected lesion visibility across anatomical site, LBR, lesion morphology, and reconstruction algorithm. Following the Rose criterion, lesions with CNR ≥ 5 were considered detectable, and the limit of detection (LOD) was defined as the smallest lesion volume meeting this threshold at the optimal post-reconstruction filter width [30]. Results are reported in Test 1 of Section 3.3.

**Detection-feature Stability:** Beyond CNR-based conspicuity, we assessed the stability of first-order image-derived features relevant to visual detection, lesion characterization, and disease staging across reconstruction and filtering settings [31,32,33,34,35]. The feature panel included SUVmax, reflecting highest focal uptake; SUV90, the 90th percentile of lesion uptake, a more noise-robust high-uptake measure; and first-order entropy, reflecting lesion intensity heterogeneity, all computed following standardized definitions [36].

For each condition, features were measured within the known inserted lesion volume and compared with ground truth. Their mean relative deviation was summarized using the Detection Stability Index (DSI):

$$\mathrm{DSI}_c = \frac{1}{|S|}\sum_{i\in S}\left|\frac{x_{i,c} - x_i^{\mathrm{GT}}}{x_i^{\mathrm{GT}}}\right| \times 100\%$$

where $S$ is the selected feature panel, $x_{i,c}$ is the measured value of feature $i$ under reconstruction condition $c$, and $x_i^{GT}$ is the corresponding ground-truth value. Lower DSI indicates better preservation of detection-relevant features. Results are reported in Test 2 of Section 3.3.

## 3. Results

The results are presented in accordance with the evaluation framework described above. We first assess the diagnostic realism of the inserted lesions, followed by task-based analyses of quantification and detection performance using the known lesion ground truth.

### 3.1 Diagnostic Realism of Inserted Lesions

The physician identified 26 of the 30 inserted synthetic lesions, while also detecting 29 native lesions, indicating that the inserted lesions had sufficient salience to be recognized under routine diagnostic reading conditions. To assess whether the two lesion classes were distinguishable, ROC analysis was performed with synthetic lesions treated as the positive class. The resulting AUC was 0.49 (95% CI: 0.34–0.64), consistent with chance-level discrimination and indicating that the observer could not reliably distinguish synthetic insertions from native pathology. This finding was supported by the overlapping confidence-score distributions for synthetic and real lesions (Table 1), validating the clinical realism of the inserted lesions and the plausibility of the proposed framework.

### 3.2 Quantification Performance Results

Having established the diagnostic realism of the inserted lesions, we next used the known ground truth to evaluate reconstruction-dependent quantitative accuracy.

***Test 1. Recovery and the Limit of Quantification:*** Recovery increased with lesion volume and iteration number for both algorithms, with lesion volume emerging as the dominant determinant of quantitative accuracy (Fig. 3). Spherical lesions showed better recovery than volume-matched patient-derived lesions because the higher surface-to-volume ratio of irregular shapes increases the fraction of voxels affected by boundary partial-volume blurring. This difference was greatest for the smallest lesions and decreased with increasing lesion volume.

Under OSEM, the largest RC gains occurred in early iterations, although recovery continued to increase gradually through iteration 10, particularly for smaller lesions. For the smallest lesion class (~0.4 mL), RC at iteration 10 was approximately 0.75–0.77 for spheres and 0.55–0.59 for patient-derived lesions, confirming substantial underestimation of small irregular lesions. For the largest class (~7.4 mL), OSEM reached RC values of approximately 0.87–0.91 for both lesion types. BSREM with $\beta = 5$ plateaued earlier and yielded lower recovery for the smallest patient-derived lesions (RC approximately 0.47–0.58 at iteration 20), while differences between OSEM and BSREM decreased with increasing lesion volume.

Penalty strength substantially affected BSREM recovery. At iteration 20, RC decreased progressively with increasing $\beta$ from 5 to 20 across lesion sizes and anatomical sites, with the greatest effect in smaller lesions (Fig. 4). In contrast, recovery of the largest lesions was less sensitive to penalty strength, remaining approximately 0.80–0.86 at $\beta = 20$. Thus, stronger regularization primarily reduced recovery of small-to-medium lesions.

To define the limit of quantification (LOQ), we applied a $|\text{Bias}| \leq 25\%$ criterion ($\text{RC} \geq 0.75$). At iteration 4, OSEM met the LOQ threshold for patient-derived lesions at ~3.7 mL and above, whereas smaller lesions remained below the criterion across all sites. At iteration 10, OSEM reduced the LOQ to ~2.0 mL across all three anatomical sites. BSREM with $\beta = 5$ achieved a similar LOQ of ~2.0 mL, while stronger regularization progressively shifted the LOQ toward larger lesions (Fig. S4).

These findings have direct implications for DLBCL PET reporting, where SUV-based metrics and metabolic tumor volume guide treatment [37,38,39]. Iteration 4 reflects a commonly used clinical setting but may provide insufficient quantitative recovery, particularly for smaller lymphoma lesions. Increasing OSEM to iteration 10 improved recovery, allowing smaller lesions to meet the predefined quantification criteria, although values near the LOQ should be interpreted cautiously. BSREM with a weak penalty ($\beta$ = 5) achieved comparable quantification limits while stabilizing earlier, whereas stronger regularization progressively increased under recovery, particularly for small lesions. Below the LOQ, lesions remained substantially underestimated regardless of algorithm, iteration number, or penalty strength.

***Test 2. Edge Effects Analysis:*** Boundary-level analysis revealed distinct reconstruction behaviors (Fig. 5). For spherical lesions under PSF-OSEM, overshoot increased monotonically from negative values at iteration 1 to approximately 40–60% by iteration 10 across all LBRs. Gibbs Activity exceeded the sharp-truth baselines (33.3%, 60.0%, 71.4%, and 77.8% for LBRs of 2:1, 4:1, 6:1, and 8:1, respectively), confirming ringing beyond the true activity step. Undershoot increased in parallel, reaching approximately 20–40% at iteration 10. For patient-derived lesions, overshoot increased from negative values at iteration 1 to approximately 15–20% by iteration 10. The lower magnitude than for spheres reflects their irregular boundary geometry, which reduces the amplitude of PSF-induced ringing.

BSREM suppressed these artifacts in a penalty-dependent manner. At $\beta$ = 5, spherical-lesion overshoot reached approximately 15–35%, while patient-derived lesions remained near zero. At $\beta$ = 20, spherical overshoot decreased to below 15%, patient-derived lesions maintained negative overshoot, and undershoot remained below 10% (Fig. S5). Gibbs Activity progressively decreased with stronger regularization.

These artefacts are clinically relevant in lymphoma response assessment, where Deauville scoring and $\Delta SUV_{max}$-based criteria rely on peak uptake. PSF-driven overshoot under OSEM increased with iteration, potentially inflating peak uptake and misclassifying metabolic response. BSREM with $\beta$ = 10 provided a favorable balance between quantitative recovery and peak-based metrics, achieving an LOQ of ~2.0 mL while suppressing edge artefacts. Stronger regularization further reduced edge effects but shifted the LOQ toward larger lesions. However, additional task-based evaluation is required to establish an optimal overall BSREM operating point.

***Test 3. The Noise–Bias Trade-off:*** The noise–bias analysis confirmed distinct behaviors between OSEM and BSREM. With OSEM, increasing iteration reduced bias from 46.8% to 23.9% while increasing noise from 21.5% to 32.8% between iterations 1 and 10 (Fig. 6). In contrast, BSREM produced shorter, penalty-dependent trajectories: at $\beta$ = 5, noise increased from 21.4% to 24.4% while bias decreased from 47.3% to 27.3%; at $\beta$ = 20, noise remained nearly constant (21.2–21.6%) while bias decreased from 48.7% to 36.1%. These findings demonstrate that BSREM suppresses noise amplification at the cost of greater residual bias.

Lesion volume was the dominant determinant of residual bias. Under OSEM at iteration 4, mean absolute bias decreased from 51.4% for the smallest real lesions to 15.4% for the largest. Spheres showed systematically lower bias than patient-derived lesions of matched volume, consistent with Test 1 findings (Fig. S6). Higher LBR consistently yielded higher bias at matched noise, most pronounced for small lesions where background spill-in at low LBR artificially inflates the measured lesion signal, reducing apparent bias.

Overall, OSEM iteration 4 provided the most favorable balance; further iterations produced progressively higher noise for comparatively smaller reductions in bias, although the higher-iteration setting resulted in a better LOQ. For BSREM, noise remained nearly stable with increasing iteration while bias continued to decrease, supporting $\beta$ = 5 at iteration 20 as the preferred operating point within the scope of this study.

### 3.3 Detection Performance Results

Detection performance was assessed using patient-derived real lesion templates of varying sizes inserted into lung, bone, and liver backgrounds at LBRs of 2:1, 4:1, 6:1 and 8:1, evaluated across two complementary tests.

***Test 1. Contrast-to-Noise Ratio and the Limit of Detection:*** CNR curves are shown in Fig. S7. Since the noise-bias analysis identified iteration 4 for OSEM and iteration 20 for BSREM ($\beta = 5$) as providing the most favorable noise-bias trade-off, the CNR and LOD analyses were plotted at these iterations. All analyses used patient-derived lesion templates with heterogeneous intra-lesional uptake, yielding more conservative LOD estimates.

The LOD decreased with increasing LBR and was comparable between OSEM and lightly regularized BSREM (Table 2). At LBR 2:1, no lesion met the detection criterion. At LBR 4:1, OSEM reached 0.70 mL in liver, 2.06 mL in bone, and 3.63 mL in lung, while at LBR 6:1–8:1 both algorithms detected 0.70 mL lesions across all sites. Strong regularization reduced detectability, requiring 3.63 mL at LBR 4:1 and preventing 0.70 mL detection in bone and liver below LBR 8:1. Filtering improved conspicuity only at LBR 2:1 (optimal FWHM 3–6 mm); at LBR 4:1 and above, CNR was highest without filtering. This likely reflects the TOF and PSF modelling incorporated in the reconstruction, which already improves signal-to-noise and recovers lesion contrast, so that additional smoothing degrades conspicuity without a compensating noise benefit.

These results establish practical detection thresholds for DLBCL PET within the scope of this study. At LBR 4:1, OSEM at iteration 4 showed site-dependent LODs ranging from 0.70 to 3.63 mL, whereas sub-mL lesions were detectable across all anatomical sites at LBR 6:1–8:1. BSREM ($\beta = 5$, iteration 20) achieved comparable or better detection, whereas stronger regularization ($\beta = 20$) reduced sensitivity. We therefore suggest OSEM at iteration 4 or BSREM with $\beta = 5$, both without post-reconstruction smoothing, reserving moderate filtering for very low-contrast lesions. These findings are relevant to disease staging, where missed small, low-contrast lesions may alter treatment selection.

***Test 2. Stability of Detection-Representing First-Order Features:*** Feature stability was evaluated using the Detection Stability Index (DSI), summarizing deviations in SUVmax, SUV90, and first-order entropy from the inserted ground truth (Fig. 7). With OSEM, DSI improved rapidly during early iterations; for the 0.34 mL lesions it decreased from 46.9% at iteration 1 to 33.5% at iteration 4 and 12.2% at iteration 20, whereas the 7.47 mL class reached a minimum of 5.1% at iteration 2 before increasing to 16.5% at iteration 20. With BSREM, DSI decreased monotonically and plateaued by approximately iteration 12 at $\beta = 5$, while $\beta = 20$ plateaued earlier at higher values. For lesions above 1 mL, OSEM iteration 4 and BSREM iteration 20 therefore provided the most favorable feature stability.

At the feature level, entropy was the most stable, remaining within 1–3% of ground truth for lesions above 1 mL. SUVmax was the least stable under OSEM, reaching its minimum at early iterations before increasing to a 35.8% deviation by iteration 20, consistent with PSF-related peak inflation, whereas SUV90 improved with convergence. Under BSREM both features stabilized early, although stronger regularization produced persistent SUV90 under-recovery. Lesion volume remained the dominant determinant of DSI: at iteration 4, deviations for the 0.34 mL class were 33.5% with OSEM and 42.5% with BSREM ($\beta = 20$), compared with 7.0% and 3.4% for the 7.47 mL class.

For lesions above 1 mL, OSEM iteration 4 therefore sufficed, since further convergence worsened SUVmax fidelity without improving recovery, whereas smaller lesions continued to benefit from higher iterations despite substantial residual partial-volume error. BSREM with light regularization ($\beta$ = 5) preserved features best overall, while stronger regularization should be used cautiously for SUVmax-based assessment of small lesions, since persistent focal uptake under-recovery could affect borderline Deauville interpretation or suppress low-contrast residual disease [40,41,42]. Reconstruction settings can therefore alter detection-relevant biomarkers without biological change, supporting size- and task-specific protocols.

## 4. Discussion

This study presents a patient-informed framework for ground-truth-based PET reconstruction optimization in DLBCL FDG imaging, where SUV-based metrics inform treatment response assessment [38,39]. By inserting synthetic lesions into clinical acquisitions, the framework preserves patient anatomy, statistical noise, and system physics while providing known ground truth for rigorous task-based evaluation beyond what physical phantoms or Monte Carlo simulations alone can offer.

Within this task-based framework, reconstruction performance depended strongly on the clinical objective rather than a single optimal setting [43]. For mean-SUV quantification, OSEM at iteration 10 reduced the quantification limit for patient-derived lesions to approximately 2.0 mL, consistent with PSF-enabled PET recovery trends [7,8], whereas iteration 4 provided a more noise-efficient setting for visual assessment and reproducibility. BSREM with weak regularization ($\beta$ = 5) achieved the same 2.0 mL limit while stabilizing earlier and suppressing PSF-related peak inflation relevant to Deauville scoring and $\Delta SUV_{max}$-based response assessment [40,41,42]. Stronger regularization increased under-recovery and shifted the quantification limit toward larger lesions. Thus, OSEM required task-specific iteration selection, whereas weakly regularized BSREM provided a more balanced compromise between recovery, noise, and peak fidelity.

Expressing recovery as a function of LBR rather than absolute uptake [12] reduced anatomical-site differences (Fig. S8), suggesting that LBR normalization reduces anatomical differences in background spill-in. For small lesions at early iterations, low-LBR spill-in artificially elevated measured activity, whereas higher LBR revealed the underlying partial-volume-limited recovery loss. This is relevant to harmonizing SUV thresholds across lesion locations [43]. Patient-derived lesions also showed lower recovery than volume-matched spheres, suggesting that phantom-based studies may overestimate quantitative reliability for clinically realistic lesion morphologies [11].

Beyond recovery, edge-artifact analysis revealed important reconstruction-dependent effects. At iteration 10, PSF-OSEM produced Gibbs-related peak overshoot of approximately 40–60% in spheres and 15–20% in patient-derived lesions, potentially affecting $SUV_{max}$, Deauville scoring, and $\Delta SUV_{max}$-based response assessment [44]. BSREM suppressed boundary artifacts in a penalty-dependent manner [29]: weak regularization reduced patient-derived overshoot to near zero, while stronger penalties further suppressed artifacts at the cost of recovery. An intermediate penalty ($\beta$ = 10) therefore provided a favorable compromise, retaining a 2.0 mL quantification limit while limiting peak distortion. For OSEM, iterations should remain limited because overshoot increased consistently with convergence. These reconstruction trade-offs were also reflected in lesion detectability.

Detection performance was driven primarily by lesion contrast and volume. Very low-contrast lesions remained below the conspicuity threshold regardless of reconstruction or filtering, confirming contrast as a limiting determinant of detectability [45]. Post-reconstruction smoothing improved conspicuity only at LBR 2:1 (optimal FWHM 3–6 mm); at higher LBRs, CNR was maximal without filtering, as smoothing reduced lesion contrast faster than background noise [46]. Overall, detection was comparable between OSEM and weakly regularized BSREM, whereas stronger regularization reduced sensitivity.

The DSI analysis extended these findings to biomarker-level stability, showing how reconstruction settings altered detection-relevant features relative to ground truth. Entropy was the most robust metric [47], remaining within 1–3% across algorithms and iterations and supporting its potential value for protocol harmonization. In contrast, $SUV_{max}$ showed the largest OSEM deviations, reaching a minimum within the first few iterations and then rising to 35.8% by iteration 20 because of PSF-related peak inflation. Under BSREM, peak metrics stabilized earlier, although stronger regularization controlled $SUV_{max}$ at the cost of persistent $SUV_{90}$ under-recovery, potentially affecting Deauville interpretation in small lesions [42,48]. Together, these findings support early OSEM stopping for lesions above 1 mL and higher iterations for smaller lesions, where partial-volume effects remain the dominant source of feature error.

The framework is broadly applicable beyond DLBCL PET. In theranostic SPECT dosimetry, it enables evaluation of lesions near high-uptake organs or complex bone structures, where attenuation, scatter, collimator response, spill-in, and partial-volume effects interact beyond conventional phantoms. In long-axial-field-of-view PET, lesion insertion can assess whether higher sensitivity enables lower activity injection, shorter scans, or delayed imaging while preserving lesion quantification and detectability. Similarly, for AI-based denoising and image enhancement, known lesion truth enables objective evaluation of whether outputs preserve lesion uptake and detectability without introducing quantitatively misleading changes.

This study had several limitations. PET evaluation was performed on a single scanner, and diagnostic realism was assessed by one physician, limiting generalizability. However, the open framework and lesion library support multi-scanner, multi-reader, and multi-tracer validation. Patient-derived libraries also depend on cohort diversity and segmentation quality. Future work will extend validation across centers, scanners, radiopharmaceuticals, and disease sites.

In summary, we developed and publicly shared LesionStudio, a patient-informed PET/SPECT lesion-insertion platform for evaluating reconstruction protocols against known lesion ground truth in clinically realistic data. By preserving patient anatomy, acquisition statistics, system physics, and realistic lesion morphology, the platform enables task-specific optimization beyond conventional phantom studies. In the lymphoma PET application, the framework showed that reconstruction choice should be matched to the clinical task: OSEM with higher iterations is preferable when mean-SUV quantification above the limit of quantification is required, OSEM with lower iterations provides a more noise-efficient setting for detection and reproducibility, and BSREM with a weak penalty achieves both at a single setting while suppressing PSF-related peak artifacts, with stronger penalties trading uptake recovery and detectability for further artifact suppression.

# Tables

**Table 1** Observer confidence-score distribution for synthetic and real lesions. Scores range from 1, definitely real, to 5, definitely synthetic.

| Confidence Score | Synthetic Lesions | Real Lesions |
|---|---|---|
| 5 | 0 | 0 |
| 4 | 6 | 8 |
| 3 | 4 | 5 |
| 2 | 8 | 6 |
| 1 | 8 | 10 |
| **Total** | **26** | **29** |

**Table 2** Limit of Detection (LOD): smallest lesion volume (mL) achieving CNR ≥ 5 (Rose criterion) at the optimal post-reconstruction filter. "—" indicates no tested volume met the criterion.

| | | **OSEM** | | **BSREM (iter 20)** | |
|---|---|---|---|---|---|
| **Site** | **LBR** | **iter 4** | **iter 8** | **β = 5** | **β = 20** |
| **Bone** | 2:1 | — | — | — | — |
| | 4:1 | 2.06 | 0.70 | 0.70 | 3.63 |
| | 6:1 | 0.70 | 0.70 | 0.70 | 2.06 |
| | 8:1 | 0.70 | 0.34 | 0.70 | 2.06 |
| **Liver** | 2:1 | — | — | — | — |
| | 4:1 | 0.70 | 0.70 | 0.70 | 3.63 |
| | 6:1 | 0.70 | 0.70 | 0.70 | 2.06 |
| | 8:1 | 0.70 | 0.70 | 0.70 | 0.70 |
| **Lung** | 2:1 | — | — | — | — |
| | 4:1 | 3.63 | 3.63 | 3.63 | 3.63 |
| | 6:1 | 0.70 | 0.70 | 0.70 | 0.70 |
| | 8:1 | 0.70 | 0.70 | 0.70 | 0.70 |

*Note*—LBR = lesion-to-background ratio; OSEM = ordered subset expectation maximization; BSREM = block sequential regularized expectation maximization; iter = iteration number; β = BSREM penalty (regularization) strength.

# Figures

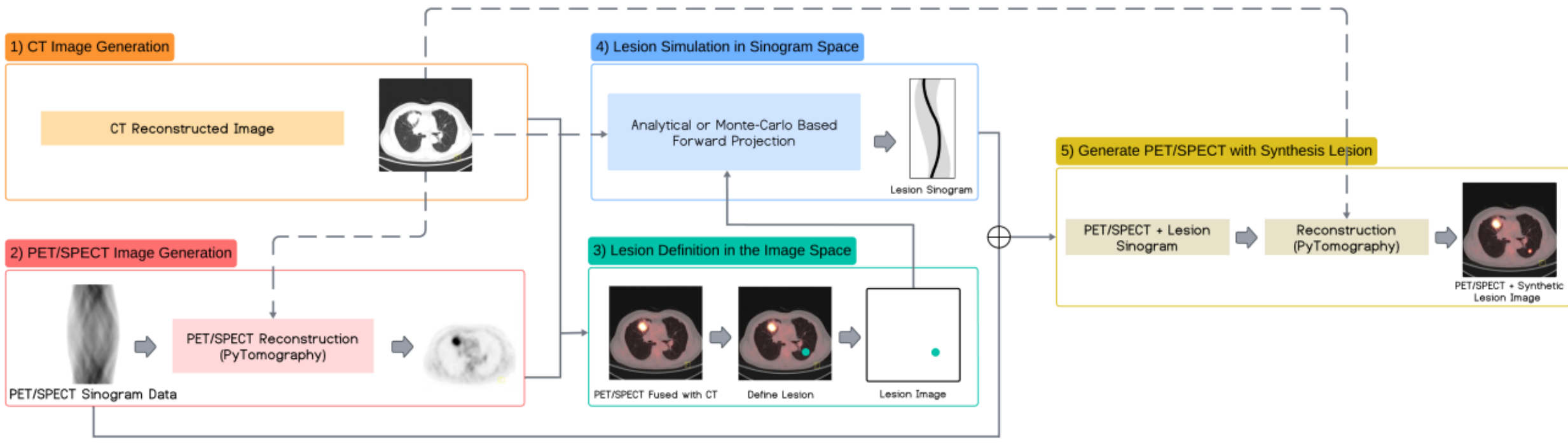

**Fig. 1**—Workflow of the proposed lesion insertion framework, illustrating CT and PET/SPECT image generation, lesion definition in image space, forward projection to sinogram or list-mode space, and reconstruction of PET/SPECT images with a synthetic lesion.

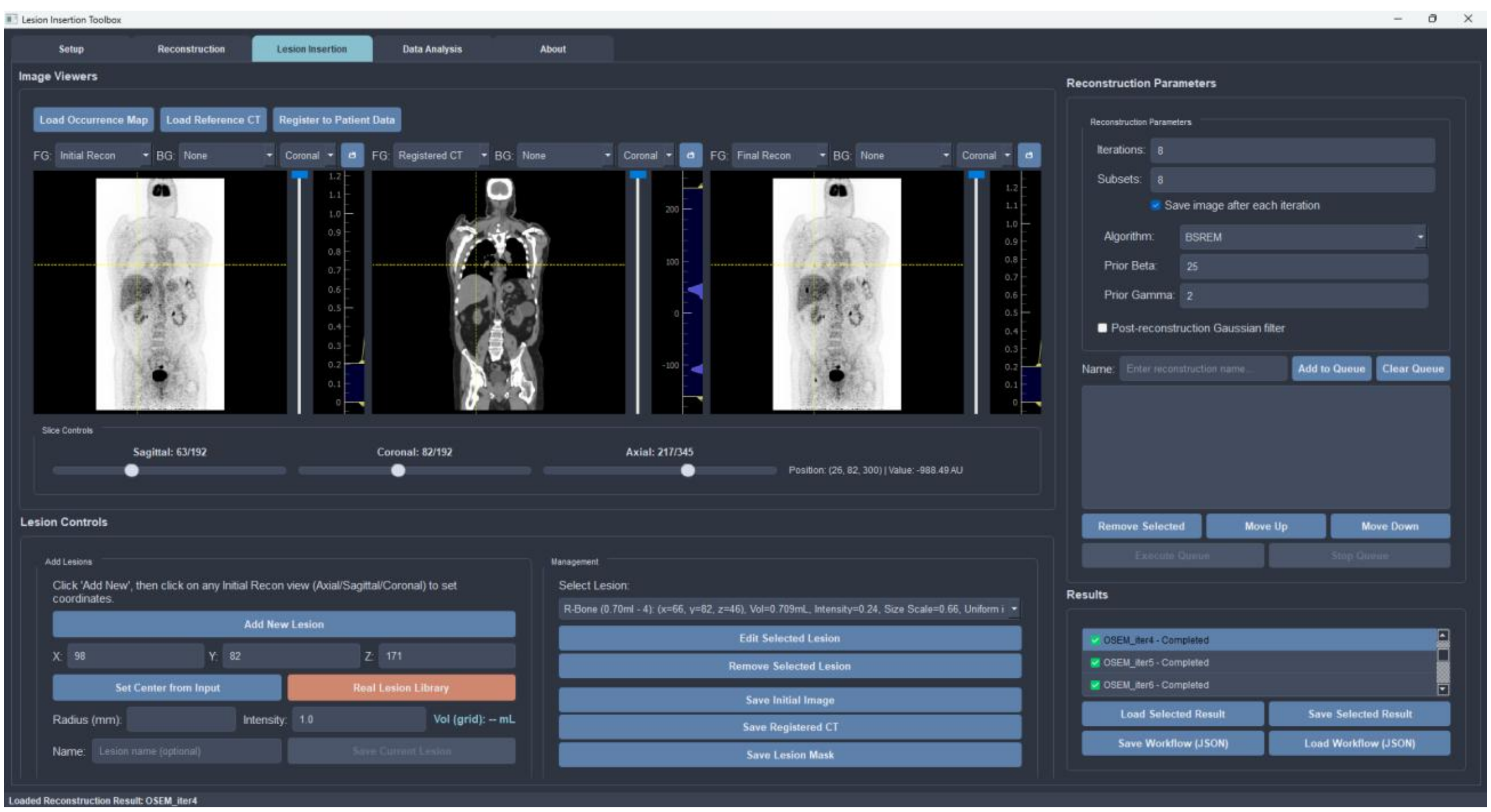

**Fig. 2**—Graphical user interface of the proposed lesion insertion framework. The interface integrates patient-data loading, fused PET/CT or SPECT/CT visualization, anatomically guided lesion placement, lesion-parameter configuration, and reconstruction control within a single workflow.

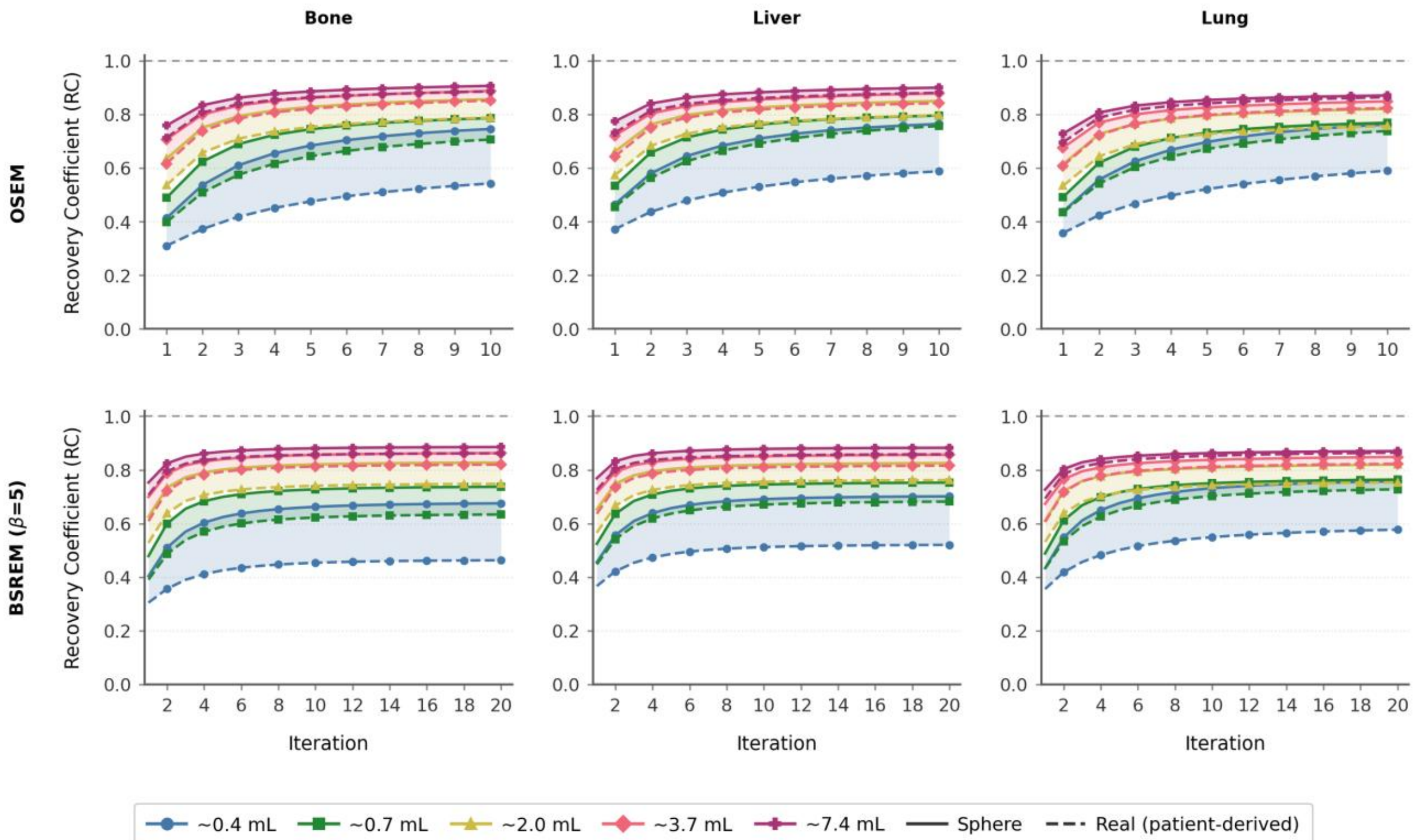

**Fig. 3—**Recovery coefficient across iterations for spherical and patient-derived lesions, stratified by reconstruction algorithm, anatomical site, and lesion volume.

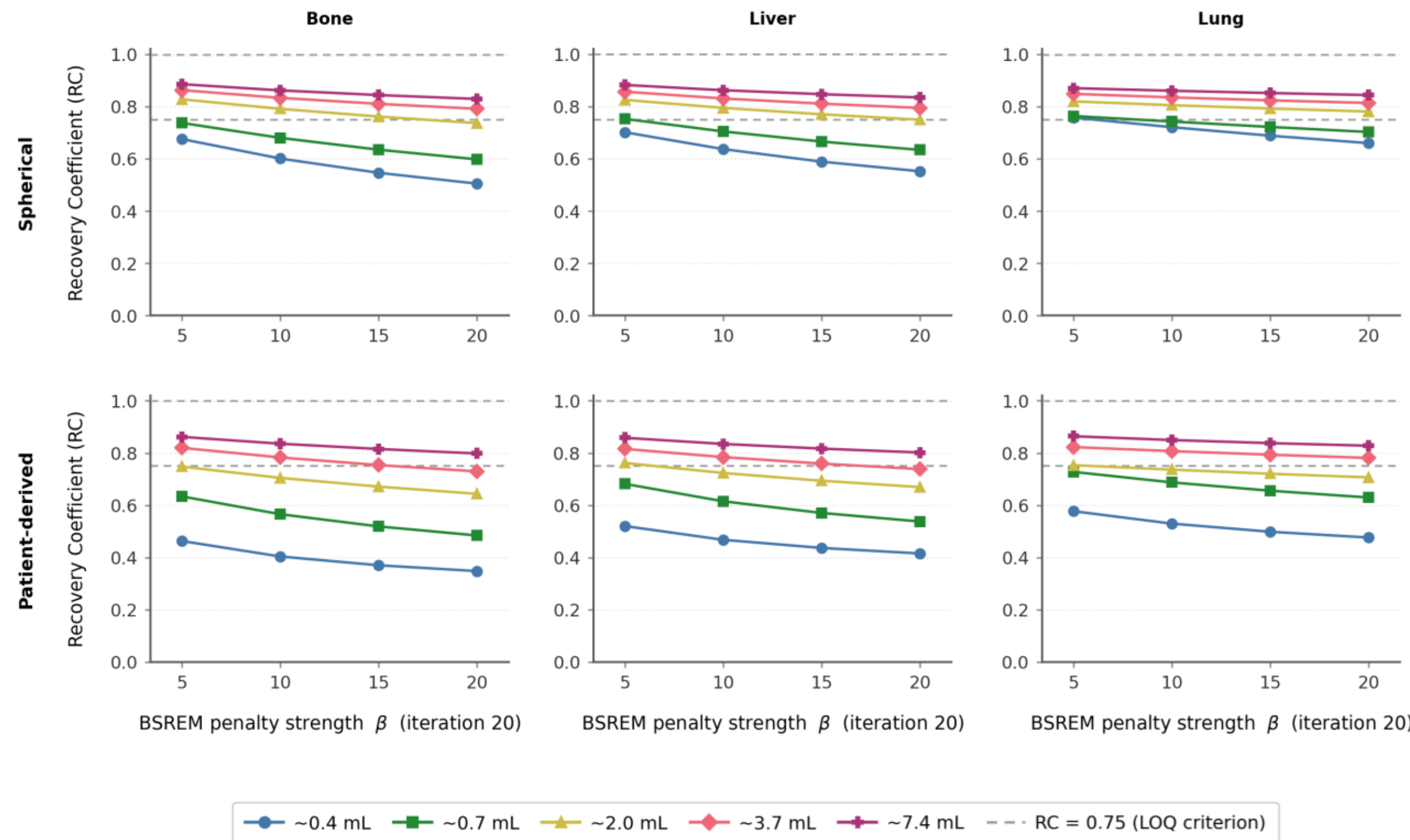

**Fig. 4—**Recovery coefficient as a function of lesion volume for spherical and patient-derived lesions reconstructed with BSREM using different penalty strengths (β = 5–20) at iteration 20 across bone, liver, and lung backgrounds.

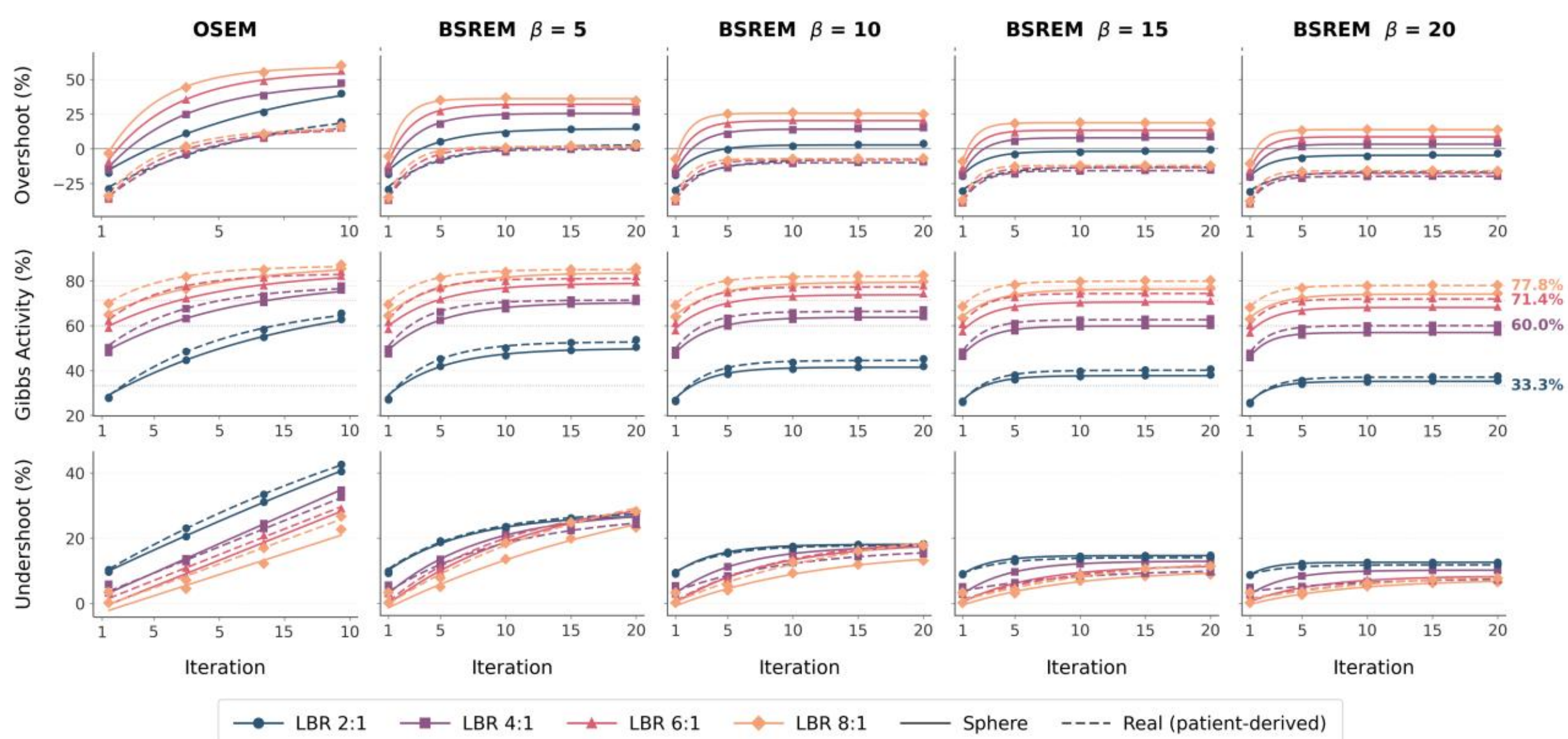

**Fig. 5—**Iteration-dependent Overshoot, Gibbs Activity, and Undershoot for spherical and patient-derived lesions reconstructed with OSEM and BSREM at β = 5–20 across LBR levels. Horizontal reference lines indicate the sharp-truth Gibbs Activity baselines for each LBR.

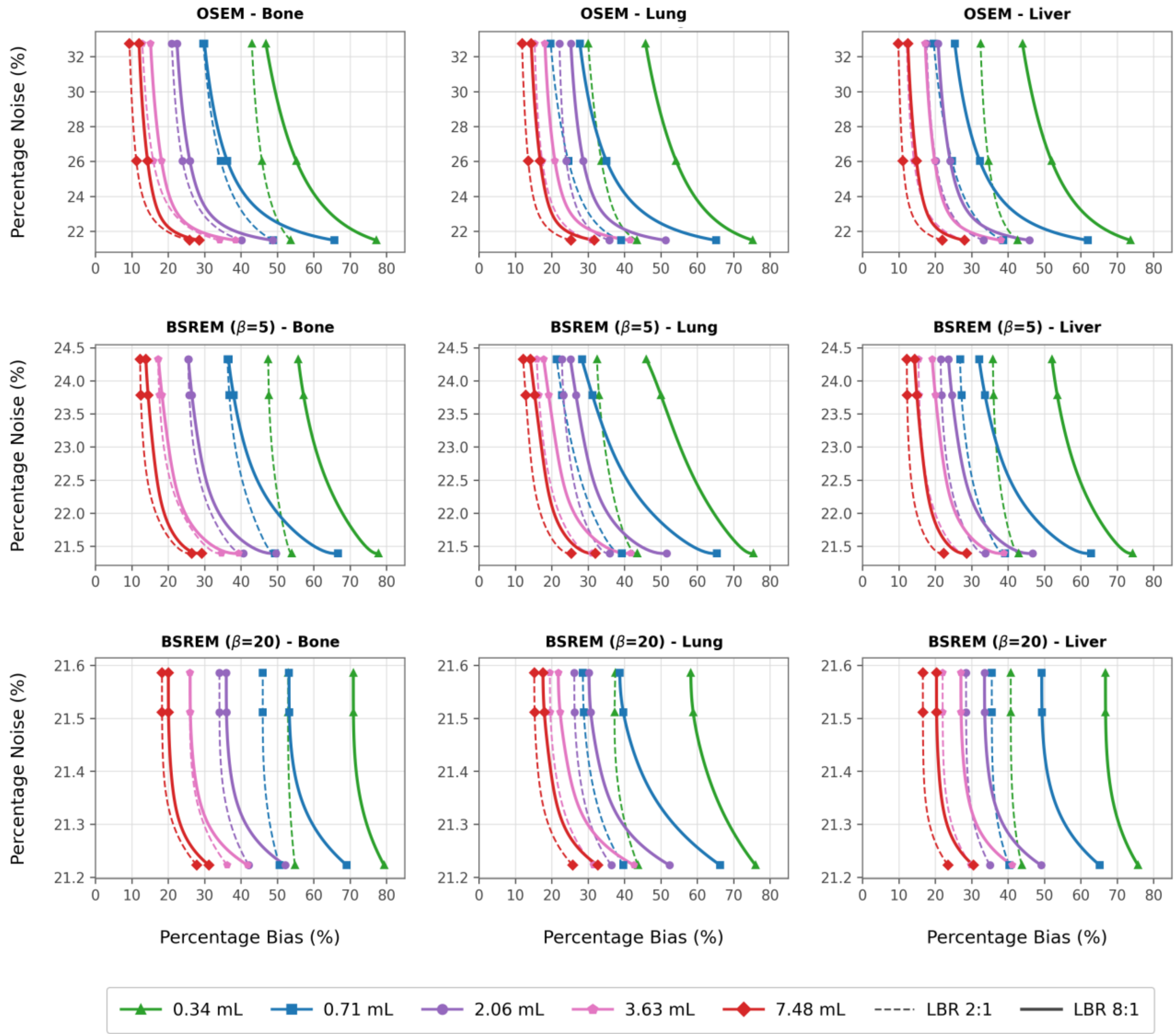

**Fig. 6—**Noise–bias trade-off curves for patient-derived lesions reconstructed with OSEM and BSREM (β = 5 and β = 20) across anatomical sites, lesion volumes, and LBR levels. Vertical scales differ between rows.

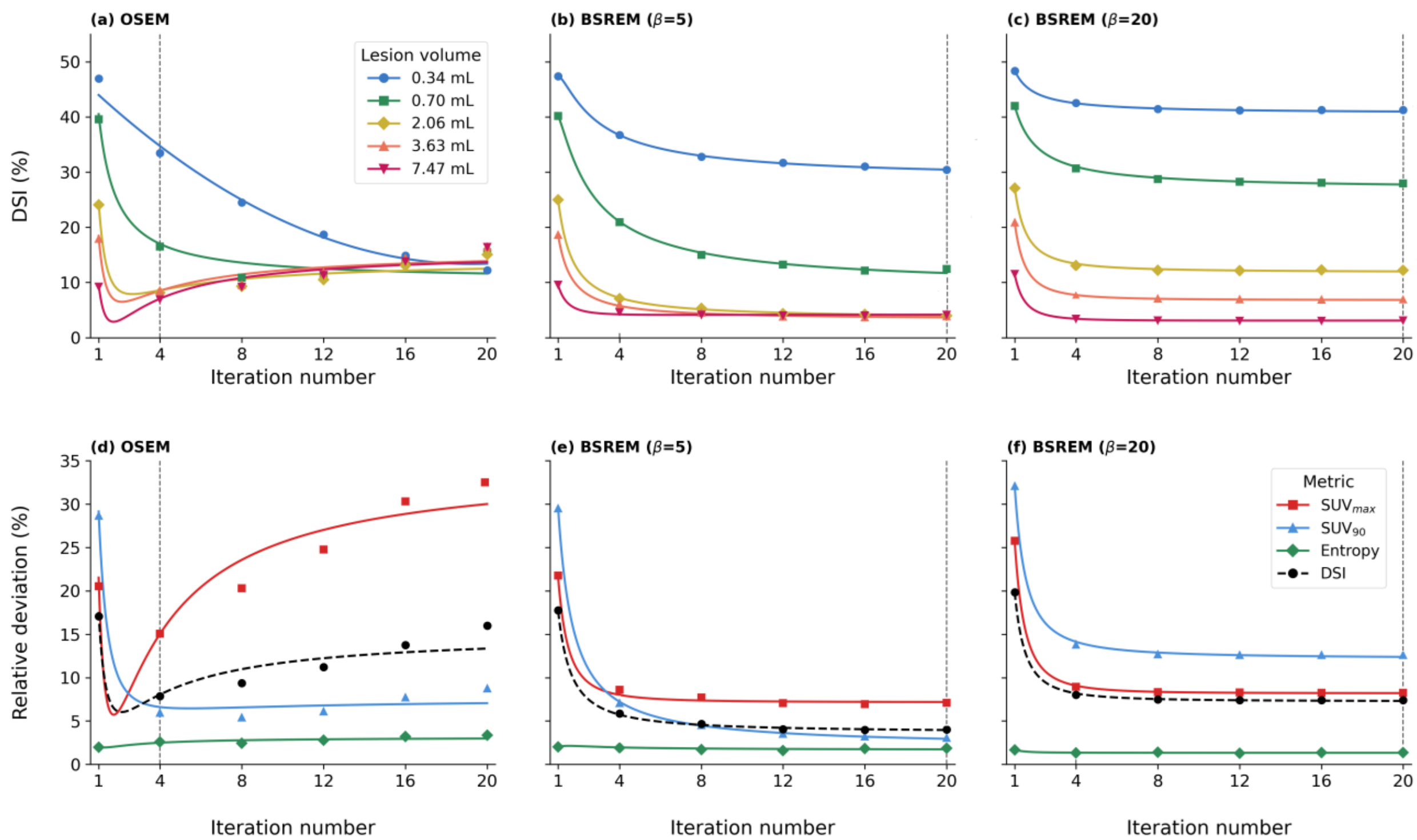

**Fig. 7—**Detection Stability Index (DSI) analysis for OSEM and BSREM, averaged across the three anatomical sites and four LBRs. (a–c) DSI across iterations by lesion volume. (d–f) Relative deviations of SUVmax, SUV90, entropy and DSI across iterations, for lesions larger than 1 mL.

## Acknowledgements

The authors thank Drs. Ran Klein and Carlos Uribe-Munoz for their valuable support and feedback, and GE HealthCare for assistance with the list-mode data processing step.

## Declarations

### *Funding*

This work was supported by the Canadian Institutes of Health Research (CIHR) Project Grant PJT-173231, and the BC Cancer Foundation.

### *Conflict of interest/Competing interests*

**Narges Aghakhanolia:** No disclosures.

**James Fowler:** No disclosures.

**Ahmad Shariftabrizi:** No disclosures.

**Omid Gharibi:** No disclosures.

**Behnam Panahi:** No disclosures.

**Sinan Akay:** No disclosures.

**Arman Rahmim:** Co-founder of Ascinta Technologies.

**Pedro L. Esquinas:** Employee of Ascinta Technologies.

### *Ethics approval and consent to participate*

This study was approved by the UBC / BC Cancer Research Ethics Board under protocol H24-03727. Retrospective imaging data were used and de-identified prior to analysis in accordance with the approved study protocol.

### *Data availability*

The lesion library is publicly available on GitHub: *https://github.com/qurit/LesionStudio*

### *Code availability*

The full framework code is publicly available on GitHub: *https://github.com/qurit/LesionStudio*

### *Generative AI use*

A generative AI-based writing assistant (Claude, Anthropic) was used solely for proofreading and grammar checking of the manuscript text. It was not used to generate, analyze, or interpret any scientific content, data, or conclusions, and all such content was produced solely by the authors.

### *Author contribution*

N.A. led the conception and design of the study, developed and implemented the lesion insertion framework, performed the reconstruction experiments and quantitative analyses, interpreted the results, and drafted the manuscript. A.S. contributed to the observer study and supported the development of the patient-derived lesion library. O.G. contributed to the creation and implementation of the real lesion library as well as radiomic feature analysis. J.F. and B.P. provided technical support for the development, validation, and implementation of the lesion insertion framework. S.A. contributed to data collection, lesion annotation, and segmentation. A.R. and P.L. supervised the project, provided conceptual, methodological, and scientific guidance, contributed to the interpretation of the findings, and critically revised the manuscript. All authors reviewed and approved the final manuscript.

# Supplementary Information: LesionStudio: A Patient-Informed Platform for Task-Specific Optimization of Emission Tomography Reconstruction and Analysis

## Supplemental Methods

### S1. SPECT Case Study: Lu-177 PSMA Prostate Cancer Dosimetry (Figures S1–S3)

To demonstrate the applicability of the proposed framework beyond PET, we present results of a case study involving Lu-177 PSMA SPECT imaging of a prostate cancer patient undergoing radiopharmaceutical therapy. We compare the quantitative accuracy of OSEM and BSREM across anatomically distinct regions (bone, lung, and abdomen) using the same synthetic lesion insertion pipeline. Synthetic lesions with diameters of 10, 20, 30, and 40 mm were inserted at two uptake levels per anatomical site. Recovery coefficients (RC), noise-bias trade-off curves, and contrast-to-noise ratios (CNR) were computed across iteration numbers from 2 to 10 at a fixed subset count of 8, and post-reconstruction Gaussian filters ranging from 1 to 10 mm FWHM were applied to assess their effect on lesion detectability.

#### *Noise–Bias Trade-off*

The noise-bias trade-off curves revealed distinct algorithmic behaviors across lesion sizes, uptake levels, and anatomical sites, as illustrated in Fig. S1. OSEM consistently achieved lower bias than BSREM, particularly for medium and large lesions, but at the cost of substantially higher noise. Conversely, BSREM suppressed noise more effectively while introducing greater bias, especially for small (10–20 mm) or low-uptake lesions. Lesion detectability improved with increasing size and uptake across both algorithms, with the lowest bias observed in the lung, intermediate in the bone, and the highest in the abdomen.

#### *Recovery Coefficients*

RC curves as a function of iteration numbers further highlighted the size- and algorithm-dependent nature of lesion recovery, as shown in Fig. S2. Small lesions (10 mm diameter) exhibited poor recovery regardless of uptake level or anatomical location, whereas larger lesions (30–40 mm) approached RC values of 0.80–0.85. Lesion recovery plateaued after 6–8 iterations for both algorithms. OSEM consistently achieved higher RC than BSREM across all lesion sizes, particularly for medium and large lesions, while BSREM plateaued earlier with lower maximum recovery. These findings underscore the importance of anatomy- and application-specific optimization: in prostate cancer dosimetry, accurate recovery of bone lesions is clinically critical, yet this cannot be achieved without accounting for the local tissue environment, a distinction that conventional phantom-based approaches cannot provide.

#### *Effect of Post-Reconstruction Filtering on CNR*

Post-reconstruction Gaussian filtering was evaluated as a means of improving lesion detectability, with results summarized in Fig. S3. An optimal filter width of 2–5 mm FWHM was identified, within which modest smoothing reduced perilesional noise without significantly compromising lesion contrast. Filters beyond this range progressively blurred lesion boundaries and reduced CNR. BSREM outperformed OSEM for small or low-uptake lesions in terms of CNR, where OSEM exhibited slower convergence and greater CNR loss with increasing filter width. Together, these results confirm that the optimal reconstruction and filtering configuration is jointly determined by lesion size, uptake level, and anatomical context, and that no single protocol setting is simultaneously optimal across the full range of clinically relevant conditions.

These preliminary SPECT findings are consistent with the PET results presented in the main body of this paper and further support the framework's utility as a modality-agnostic platform for task-specific reconstruction optimization. A comprehensive multi-patient SPECT evaluation, including Lu-177 PSMA and additional radiopharmaceuticals, is planned as part of future work.

## S2. Additional PET Reconstruction Analyses (Figures S4–S8)

Figures S4–S8 provide additional quantification and detection analyses for the FDG PET dataset described in Section 2.2 of the main text, complementing the recovery, edge-artifact, noise–bias, and detectability results reported in Sections 3.2 and 3.3. Fig. S4 extends the limit-of-quantification analysis of Section 3.2 (Test 1) by comparing OSEM at iteration 10 with BSREM at iteration 20 directly across anatomical sites. Fig. S5 summarizes the edge-artifact analysis of Section 3.2 (Test 2) as a heatmap across all tested lesion types, volumes, LBRs, algorithms, penalty strengths, and iterations. Fig. S6 reports the noise–bias trade-off for spherical lesions (Section 3.2, Test 3), complementing the patient-derived results shown in Fig. 6 of the main text. Fig. S7 shows the contrast-to-noise ratio analysis underlying the limit-of-detection results in Section 3.3 (Test 1). Fig. S8 expresses recovery coefficient as a function of lesion-to-background ratio, referenced in the Discussion.